\documentstyle[preprint,tighten,prd,aps,eqsecnum]{revtex}
\def\eval#1{\left\langle#1\right\rangle}
\def\bA{{\bf A}}
\def\bm{{\bf m}}
\def\bn{{\bf n}}
\def\be{{\bf e}}
\def\br{{\bf r}}
\def\bs{{\bf s}}
\def\bk{{\bf k}}
\def\bx{{\bf x}}
\def\by{{\bf y}}
\def\bz{{\bf z}}

\begin{document}
\draft
\preprint{OUTP-98-75 P}
\title{Interaction potential in compact three-dimensional QED with mixed action}
\author{A. Kovner}
\address{Theoretical Physics, University of Oxford,
1 Keble Road, Oxford, OX1 3NP, United Kingdom}
\author{B. Svetitsky}
\address{School of Physics and Astronomy, Raymond and Beverly Sackler Faculty
of Exact Sciences, Tel~Aviv University, 69978 Tel~Aviv, Israel}
\maketitle

\begin{abstract} 
We use a variational wave function
to calculate the energy of the interaction between
external charges in the compact Abelian gauge theory in 2+1 dimensions with
mixed action.
Our variational wave functions preserve
the compact gauge invariance of the theory both in the vacuum and in the
charged sectors. 
We find that a good estimate of the interaction energy is obtained only
when we allow more variational parameters in the charged sector than
in the vacuum sector. These extra parameters are the profile
of an induced electric field.
We find that the theory has a two-phase structure: 
When the charge-2 coupling is large and negative
there is no mass gap in the theory and no confinement,
while otherwise a mass gap is generated
dynamically and the theory confines charges.
The pure Wilson theory is in the confining phase.
\end{abstract}


\section{Introduction}

The Abelian gauge theory in three dimensions has long been considered the 
archetype of a confining field theory.
Polyakov \cite{polyakov}
demonstrated charge confinement with a classic analysis of the instanton plasma;
the hope that the result would carry over to the instantons of Yang-Mills
theory in four dimensions proved, unfortunately, to be unfounded \cite{CDG}.
In the hands of Banks, Myerson, and Kogut \cite{BMK}, the theory became 
an exemplar of lattice gauge theory, displaying a single, confining phase
from the trivial limit of strong coupling down to the expected essential
singularity at weak coupling.\footnote{
The corresponding demonstration for the four-dimensional non-Abelian theory
still depends on numerical simulation \cite{Creutz}.}
It was also a stepping stone to elucidation of the four-dimensional Abelian 
theory, which confines at strong coupling but contains only weakly
interacting photons at weak coupling:
The difference between the two phases is due to a condensate of magnetic
monopoles at strong couping, closely related to the instantons of the
three-dimensional theory.
Subsequently, the Hamiltonian version of the theory became a laboratory for 
development of variational methods \cite{DQSW}, providing in this formalism 
as well a specimen of nonperturbative confinement persisting down to weak
coupling.

In recent years there has been a revival of interest in 
variational approaches to 
{\em non-Abelian} gauge theories.
Several calculations 
based on different variational {\em ans\"atze} have appeared in the literature
\cite{koko1,iancu}.
The approach of \cite{koko1}, a gauge invariant generalization
of the Gaussian (mean field) variational approximation,
was applied to the Abelian theory in \cite{KK}, where it is clear that
the wave function is a generalization of that used in \cite{DQSW} long ago.
In the Yang Mills theory \cite{koko1},
there emerges a nonperturbative
infrared mass scale, which makes
the behavior of low momentum gluon
modes in the vacuum very different from that calculated in perturbation theory.
The comparison of the ultraviolet properties of this variational wave function
with weak coupling perturbation theory
was extended in \cite{brko,brown}.

The results presented in \cite{koko1,brko,brown} are 
limited to the wave function of the vacuum state.
The most interesting phenomenon in
non-Abelian gauge theories, however,
is confinement of fundamental charges.
In a Lorentz invariant 
theory one could, in principle, calculate instead the vacuum average of a 
spacelike Wilson loop. 
In the Hamiltonian framework, however, Lorentz invariance is not explicit
and most variational approaches will preserve it (at best) only approximately.
The relation in the variational framework between the vacuum average 
of the spacelike Wilson loop and the 
energy of external charges is poorly understood.
It is therefore all the more interesting to set up a 
variational calculation for the charged sector.

In some recent papers \cite{zare,dyak}
it was suggested that the charged sector can be described
by the vacuum wave function, modified only to satisfy the new
Gauss' law constraint.
It was conjectured
that further modification of the wave 
function in the presence of the external charges is unnecessary.
We will refer to this state as the minimally modified vacuum. 
(Its exact definition will be given in Section 3.)
It is far from evident that this simple procedure gives a good
estimate of the energy of external charges. 
It does of course give a variational 
upper bound, but the quality of this bound may be poor.
External charges introduce via Gauss' law an electric flux
that is not present in the vacuum.
It may be necessary to introduce additional variational parameters to allow 
this flux to spread out in the way that is most favorable energetically.

The Abelian gauge theory in 2+1 dimensions is an ideal theater in which to
examine this issue.
The authors of \cite{DQSW} calculated  the energy of an external dipole in this
theory with the prescription advocated in \cite{zare,dyak}.
The result was
that the energy of the dipole is 
indeed proportional to its length and the string tension was determined.
The vacuum expectation value
of the spacelike Wilson loop was calculated in the same 
approach in \cite{KK}.
There it was found that the Wilson loop obeys an area law with a
string tension consistent with the result of Polyakov \cite{polyakov}.

Superficially the results of \cite{DQSW} and \cite{KK} appear consistent.
A closer look, however, reveals important differences. 
First, the vacuum state of \cite{KK} incorporates a dynamically generated
mass whereas the correlations of magnetic and electric fields in
the vacuum state of \cite{DQSW} have power law decays.
Second, the string tension calculated in \cite{DQSW} (from the energy of 
external charges) is {\em parametrically}
smaller than the result of \cite{KK} (from the spacelike Wilson loop).
Finally, a calculation of the spacelike Wilson loop with the wave function of
\cite{DQSW} gives no area law.
In the present paper we explain these discrepancies 
while providing a more general discussion of the calculation of
the energy in the presence of external charges.

We also take this opportunity to extend the variational analysis to the
theory with mixed action.
We define the theory with the lattice Hamiltonian
\begin{equation}
H=\frac12a^2\sum E^2_{\bn i}-\frac1{g^2a^2}\sum
\left(\alpha\cos ga^2B_{\bn} +\frac{\beta}4\cos2ga^2B_{\bn}\right)
\label{ham}
\end{equation}
Here $a$ is the lattice spacing and the sums are respectively over the links
and plaquettes of the two-dimensional spatial lattice. 
We constrain the constants $\alpha$ and $\beta$ to satisfy 
\begin{equation}
\alpha+\beta=1
\end{equation}
so that in the weak coupling limit, upon formal expansion to lowest
order in $g^2$, the Hamiltonian reduces to the standard free
Hamiltonian of 2+1 dimensional electrodynamics.
We work in this weakly coupled regime, $g^2a\ll1$, but
go beyond perturbation theory with a variational wave function.

The mixed Hamiltonian provides a new parameter for the theory.
It has long been believed that 
there is no non-confining phase in three dimensions,
but generalized couplings along the lines of (\ref{ham}) have not been
adequately explored.
Here we discover that this simplest of confining theories yet contains
a surprise: a second-order confinement-deconfinement 
phase transition in the parameter $\alpha$. 
We present in Section 2 a variational calculation that
determines the best variational vacuum of the theory. 
For $\alpha<{\pi^2\over 4}$ the energy is minimized for a state with a 
nonvanishing ``mass gap,'' meaning that the correlation functions of
the electromagnetic field decay exponentially at large
distances. We will refer to this phase as {\em massive}. 
(This vacuum state
is the same as that discussed in \cite{KK}.)
For $\alpha>{\pi^2\over 4}$ no such mass gap is generated, and
the correlation functions at large distances have a power law decay. 
This phase will be called {\em massless\/} in the following. 
(This state is the same as that considered in \cite{DQSW}.)
Note that the Wilson action $\alpha=1$ is in the massive phase.\footnote{
Morris \cite{Morris} has studied the three-dimensional Abelian theory
with generalized action and found phase transitions to phases with
non-zero magnetic field.
We have fixed $\alpha+\beta=1$ to prevent this in the weak-coupling limit
we study. Our new, non-confining phase is unrelated to Morris'.}

We also calculate the expectation values of Wilson loops in the two
phases. In the massive phase we find an area law for Wilson loops of
arbitrary charge, with the string tension proportional to the square of
the charge. In the massless phase the 
string tension vanishes and there is no area law.

In Section 3 we calculate the potential between two external charges. 
First we consider the minimally modified vacuum according to the recipe of
\cite{DQSW,zare,dyak}.
We calculate the profile of the electric field
and the energy of a dipole configuration. 
We find that the
results of this calculation are {\em inconsistent} with what we would infer 
from the behavior of the spacelike Wilson loop.
In the massive phase, where we found an area law,
the electric field does not form a flux tube but rather
takes the form of an ordinary, unconfined dipole field.
(This is so even when the distance between the charges is much larger than the 
dynamical mass in the vacuum.)
In the massless phase, amusingly,
the situation is reversed. Even though there is no area law, 
the field far from the charges
is screened and decays faster than a dipole field (but still as a power).

The calculation of the energy gives the following results, again 
inconsistent with the results of Section 2. 
In the massless phase we reproduce the result of \cite{DQSW},
a linear confining potential.
In the massive phase 
the energy of the minimally modified state diverges logarithmically in 
the infrared. 

These results convince us that the minimally modified state 
provides a poor upper bound for the energy of the external dipole.\footnote{
Ben-Menahem \cite{Shahar} noted that the massless wave functions of
\cite{DQSW} gave correlations inconsistent with the Euclidean analysis
\cite{polyakov,BMK},
and proposed a massive propagator to correct the discrepancy.}

In Section 4 therefore we introduce additional variational parameters
into the trial state. 
These parameters are natural in the framework of the Gaussian
approximation and correspond to a classical electric field present in the
variational state. 
Varying the energy with respect to this classical field,
we recover the results expected on the basis of the calculation in Section 2.
In the massive phase the best variational state has the structure of an
electric flux tube.
Its energy is infrared finite and proportional to the distance between the charges.
The string tension is the same as that calculated in Section 2.
In the massless phase the flux spreads out into a dipole field, and 
correspondingly the energy has no linear dependence on the dipole size. 

Finally in Section 5 we discuss our results and their implications for a
possible calculation of the string tension in non-Abelian gauge theories.

\section{The variational vacuum}
\label{sec2}
\subsection{The wave function}

For a weakly coupled theory we expect the vacuum wave function
to be not too different from that of a free theory. 
We therefore use a gauge
invariant projection of a Gaussian wave function, as discussed in detail
in \cite{KK}.

We present the formalism here in lattice notation, corresponding to the
Hamiltonian (\ref{ham}).
(We suppress the lattice spacing $a$ henceforth.)
Define a vortex field $A^V_{\bn i}$ that satisfies
\begin{equation}
\left(\nabla\times\bA^V\right)_{\bn'}=\frac{2\pi}g\delta_{{\bn'},0}
\ ,\qquad
\nabla\cdot\bA^V=0\ ,
\label{vortex1}
\end{equation}
where $\bn'$ is a site of the dual lattice.
This is the vector potential corresponding to a magnetic field that is zero
everywhere except at $\bn'=0$, where it takes the value $\frac{2\pi}g$.
We demand that our variational wave function $\psi[\bA]$
be invariant under shifts $\bA\to\bA+\bA^V$, which is an expression of the
periodicity of $H$ under $B\to B+\frac{2\pi}g$.
We also demand gauge invariance of the wave function.
Hence we define a shifted field, shifted by a gauge transformation $\phi_\bn$
and by a vortex distribution $m_{\bn'}$,
\begin{equation}
\bA^{(\phi,m)}_\bn=\bA_\bn-(\nabla\phi)_\bn
-\sum_{\bn'}m_{\bn'}\bA^V(\bn-{\bn'})
\label{shift1}
\end{equation}
or, for short,
\begin{equation}
A^{(\phi,m)}=A-\nabla\phi-A^V\cdot m\ .
\label{shift2}
\end{equation}
Then the gauge invariant and periodic trial wave function is
\begin{equation}
\psi[\bA]=\sum_{\{m_{\bn'}\}}\int[d\phi_\bn]\,
\exp\left[-\frac12\sum_{\br,\bs}A^{(\phi,m)}_{\br i}
G^{-1}(\br-\bs)A^{(\phi,m)}_{\bs i}\right]\ .
\label{wf}
\end{equation}
Under a gauge transformation,
\begin{equation}
\psi[\bA+\nabla\lambda]=\psi[\bA]
\label{g_invariance}
\end{equation}
since $\lambda$ can be absorbed in a shift in $\phi$.
The simple rotational structure of $G_{ij}=\delta_{ij}G$ that appears in the
variational wave function (\ref{wf})
is consistent with perturbation theory, as discussed in \cite{koko1}. 
In this paper we take $G(x)$ to be a real function.
We now proceed to calculation of expectation values in
the trial state (\ref{wf}) and the minimization of the 
vacuum energy.

\subsection{Normalization integral}
The normalization of $|\psi\rangle$ is
\begin{equation}
Z\equiv\langle\psi|\psi\rangle=\sum_{\{m,m'\}}\int[d\phi][d\phi'][d\bA]\,
e^{-\frac12A^{(\phi,m)}G^{-1}A^{(\phi,m)}}
e^{-\frac12A^{(\phi',m')}G^{-1}A^{(\phi',m')}}\ .
\end{equation}
We shift \bA\ by $\nabla\phi'+\bA^V\cdot m'$, and absorb the shift into
$\phi$ and $m$, giving
\begin{equation}
Z=\sum_{\{m\}}\int[d\phi][d\bA]\,
e^{-\frac12A^{(\phi,m)}G^{-1}A^{(\phi,m)}}
e^{-\frac12AG^{-1}A}\ .
\label{Z}
\end{equation}
Now we combine the exponents according to
\begin{equation}
A^{(\phi,m)}G^{-1}A^{(\phi,m)}+AG^{-1}A=
2A^{(\phi/2, m/2)}G^{-1}A^{(\phi/2, m/2)}
+\frac12 S(\phi,m)G^{-1}S(\phi,m)
\end{equation}
where
\begin{equation}
S\equiv\nabla\phi+A^V\cdot m\ .
\end{equation}
The first term in {\bf S} has zero curl while the second is divergenceless.
Furthermore, $G^{-1}$ is translation invariant and proportional to the
unit matrix.  Thus $SG^{-1}S$ has no cross terms between $m$ and $\phi$.
We now shift \bA\ by $\nabla\phi/2+\bA^V\cdot m/2$, and all the fields
decouple.
We have then
\begin{equation}
Z=Z_A Z_{\phi} Z_v\ ,
\end{equation}
where
\begin{equation}
Z_A=\det \pi G\ ,
\end{equation}
\begin{eqnarray}
Z_{\phi} &=&\int[d\phi]e^{-\frac14\nabla\phi\cdot G^{-1}\cdot\nabla\phi}\cr
&=&\left(\det 4\pi\frac1{\nabla^2}G\right)^{1/2}\ ,
\end{eqnarray}
\begin{equation}
Z_v=\sum_{\{m_{\bn'}\}}\exp\left[-\frac1{4g^2}\sum_{\br',\bs'}m_{\br'}
D({\br'}-{\bs'})m_{\bs'}
\right]\ .
\label{Zv1}
\end{equation}
$Z_v$ is the vortex partition function,
with the vortex--vortex interaction $D$ given by
\begin{equation}
D({\br'}-{\bs'})=g^2\sum_{\br,\bs}\bA^V({\br}-{\br'})\cdot
G^{-1}(\br-\bs) \bA^V(\bs-\bs')\ .
\end{equation}
The explicit solution of (\ref{vortex1}) is
\begin{equation}
A^V_i(\bn)=-\frac{2\pi}g\epsilon_{ij}
\left(\frac{\partial_j}{\nabla^2}\delta_{\bn',0}
\right)_{\bn}
\end{equation}
so
\begin{equation}
D=-g^2A^VG^{-1}A^V=-4\pi^2\frac1{\nabla^2}G^{-1}\ .
\label{DG1}
\end{equation}
We can split off the $\br'=\bs'$ terms in (\ref{Zv1}) and write
\begin{equation}
Z_v=\sum_{\{m_{\bn'}\}}\exp\left[-\frac1{4g^2}\sum_{\br'\not=\bs'}m_{\br'}
D({\br'}-{\bs'})m_{\bs'} \right]
\prod_{\br'}z^{m_{\br'}^2}
\label{Zv2}
\end{equation}
where we have defined the vortex fugacity
\begin{equation}
z=e^{-\frac1{4g^2}D(0)}\ .
\label{fugacity}
\end{equation}

In the interest of clarity, we adopt henceforth a continuum notation, indicating
the ultraviolet cutoff $a$ only where necessary.
Moreover, since the interesting physics comes from the infrared, lattice
effects can be approximated by a
momentum-space cutoff $\Lambda=a^{-1}$, which can simplify formulas further.

The variational function $G$ appears both in the vortex--vortex potential and
in the vortex fugacity.
We expect the UV behavior of $G$ to be the same as in the free theory,
{\em viz.},
\begin{equation}
G^{-1}(k)\sim k\ ,
\end{equation}
so
\begin{eqnarray}
D(0)&\sim&\int^\Lambda\frac{d^2k}{(2\pi)^2}\frac{4\pi^2}{k^2}G^{-1}(k)
\label{DG}\\
&\sim&2\pi\Lambda
\end{eqnarray}
and thus
\begin{equation}
z\sim e^{-\frac{\pi}2\frac1{g^2a}}
\label{potential}
\end{equation}
In the weak coupling region we have $z\ll1$, justifying a restriction to
$m=0,\pm1$ in  (\ref{Zv1}) and (\ref{Zv2}).

\subsection{Expectation values}

We calculate correlation functions of $m$
as in \cite{KK} via a duality transformation.
We add an $iJ\cdot m$ term to the exponent in (\ref{Zv1})
and use the formula
\begin{equation}
e^{-\frac1{4g^2} m\cdot D\cdot m}
=\text{\it const}\int[d\chi]\, e^{-g^2\chi\cdot D^{-1}\cdot\chi}
e^{i\chi\cdot m}
\label{duality}
\end{equation}
to obtain 
\begin{equation}
Z_v=\int[d\chi]\, e^{-g^2\chi\cdot D^{-1}\cdot\chi}
\prod_{\bn} [1+2\cos(\chi_{\bn}+J_{\bn})]
\label{sin1}
\end{equation}
Noting that\footnote{The normal 
ordering is performed relative to the free 
theory defined by the quadratic action in (\ref{sin1}).}
\begin{equation}
\cos (\chi+J)=\eval{\cos\chi}_0:\!\cos (\chi+J)\!:\,=z:\!\cos (\chi+J)\!:
\label{normalorder}
\end{equation}
we have
\begin{eqnarray}
Z_v&=&\int[d\chi]\, e^{-g^2\chi\cdot D^{-1}\cdot\chi}
\prod \left[1+2z:\!\cos (\chi+J)\!:\right]\nonumber\\
&\simeq&
\int D\chi\exp\left[-g^2\chi D^{-1} \chi+
2z\int d^2x\,:\!\cos\Big(\chi(x)+J(x)\Big)\!:\right]\ ,
\label{sine}
\end{eqnarray}
in continuum notation.
Correspondingly \cite{KK},
\begin{equation}
\eval{m(x)m(y)}=2g^2D^{-1}(x-y)-4g^4\eval{D^{-1}\chi(x)\,D^{-1}\chi(y)}\ .
\end{equation}

The propagator of $\chi$ is easily calculated. 
To first order in $z$,
the only contribution comes from the tadpole diagrams, which
have already been subtracted in (\ref{sine}).
Therefore
\begin{eqnarray}
\int d^2x\,e^{ikx}\eval{\chi(x)\chi(0)}&=& \frac{1}{2g^2D^{-1}(k)+2z}\nonumber\\
&=&\frac{D(k)}{2g^2} -z\frac{D^2(k)}{2g^4} + O(z^2)\ .
\end{eqnarray}
The correlation function of the vortex density is then
\begin{equation}
K(k)\equiv\int d^2x\,e^{ikx}\eval{m(x)m(0)}=2z + O(z^2)\ ,
\label{rhoc}
\end{equation}
which in this approximation does not depend on momentum.
The $k$ dependence will appear in the higher orders in $z$.

Now we are ready to calculate 
the expectation value of the Hamiltonian (\ref{ham}).
The calculation of the electric part of the energy is identical to
that in \cite{KK}.
Using the definition (\ref{wf}) we calculate
\begin{eqnarray}
V^{-1}\eval{\int E^2\,d^2x}&=&-\frac1V\eval{\psi\left|\sum_{\bn,i}
\frac{\partial^2}{\partial A_{\bn,i}^2}\right|\psi}\nonumber\\
&=&\frac12\int \frac{d^2k}{(2\pi)^{2}} G^{-1}(k)-\frac{\pi^2}{g^2}
\int \frac{d^2k}{(2\pi)^{2}}  k^{-2}G^{-2}(k)K(k)\label{elen1}\\
&=&\frac{1}{2}\int \frac{d^2k}{(2\pi)^{2}}  G^{-1}(k)-\frac{2\pi^2}{g^2}
z\int \frac{d^2k}{(2\pi)^{2}}  k^{-2}G^{-2}(k)
\label{elen}
\end{eqnarray}
The magnetic part is easily calculated since it has an exponential
form and therefore with our trial wave function 
leads to a simple Gaussian integral.
We find
\begin{equation}
\eval{e^{ingB_{\bn}}}=\exp\left[-{1\over 4}n^2g^2\int{d^2k\over
(2\pi)^2}k^2G(k)\right]\eval{e^{in\pi m_{\bn}}}\ .
\end{equation}
The second factor is due to the vortices, and is
different from unity only for odd values of $n$.
Using (\ref{sine}) we find easily that $\eval{e^{i\pi m}}=e^{-4z}$.
Expanding to leading order in $g^2$ and $z$,
\begin{equation}
\eval{-\frac1{g^2}\left(\alpha \cos gB+\frac{1-\alpha}4\cos 2gB\right)} =
\frac14\int {d^2k\over (2\pi)^2}
k^2G(k)+{4\alpha\over g^2}z\ .
\label{Benergy}
\end{equation}
(We have dropped an additive constant.)
Finally, the expression for the variational vacuum expectation value
of the energy is
\begin{equation}
\frac{1}{V}\eval H= \frac{1}{4} \int\frac{ d^2k}{(2\pi)^2}
\left[G^{-1}(k)+k^2G(k) -\frac{4\pi^2}{g^2} z\left(k^{-2}G^{-2}(k)-
{4\over \pi^2} \alpha\right) \right]\ .
\label{eval}
\end{equation}

\subsection{Determination of the ground state}

The expression (\ref{eval}) has to be minimized with  respect to $G$.
 From equation (\ref{fugacity}) and (\ref{DG})  we find
\begin{equation}
\frac{\delta z}{\delta G(k)}=\frac{1}{4g^2}k^{-2}G^{-2}(k)\,z
\end{equation}
The variation of (\ref{eval}) gives\footnote{We have dropped a term 
$-\frac{8\pi^2}{g^2}zk^{-2}G^{-3}(k)$ from the right
hand side of (\ref{variation})  is smaller by a factor of
$\frac{g^2k}{\Lambda^2}$ than the term retained, assuming $G\sim k^{-1}$
at large $k$.}
\begin{equation}
{k^2-G^{-2}(k) = 
\frac{4\pi^4}{g^{4}} 
z k^{-2} G^{-2}(k) 
\int \frac{d^2p}{(2\pi)^2} \left[p^{-2}G^{-2}(p)- {4\over \pi^2}
\alpha\right]}\ .
\label{variation}
\end{equation}
Eq.~(\ref{variation}) has the solution
\begin{equation}
G^{-2}(k)=\frac{k^4}{k^2+m^2}
\label{solution}
\end{equation}
where
\begin{equation}
{m^2=
\frac{4\pi^4}{g^4}
z\int \frac{d^2k}{(2\pi)^2} \left[k^{-2}G^{-2}(k)- {4\over\pi^2} \alpha\right]}
\ .
\label{mass}
\end{equation}
The main contribution to the integral 
in the gap equation (\ref{mass}) comes from momenta $k^2\gg m^2$. For these momenta $k^2G^{-2}(k)=1$. We 
therefore see that (\ref{mass})
has a nontrivial solution when $\alpha<{\pi^2\over 4}$.
Using equations (\ref{fugacity}), (\ref{DG}),  and (\ref{solution}) we obtain
\begin{equation}
{m^2=
\frac{4\pi^4}{g^4}
\exp\left(-\frac{\pi^{2}}{g^{2}}
\int \frac{d^2p}{(2\pi)^2} \frac{1}{\sqrt{p^2 + m^{2}}}\right) 
\int \frac{d^2k}{(2\pi)^2} 
\left[\frac{k^2}{k^2+m^2}-{4\over\pi^2}\alpha\right]}
\label{complicatedm}
\end{equation}
which for $g^2\ll1$ can be simplified to [cf.~(\ref{potential})]
\begin{equation}
{m^2 = 
{4\pi^2}
\frac{(\pi^2-4\alpha)}{g^4}
\exp\left(-\frac{\pi}{2g^2}\right)}\ .
\end{equation}
$m$ is the mass gap of the theory, in the sense that it is the
inverse of the spatial correlation length. Calculating,
for example, the propagator of magnetic field, we find
\begin{equation}
\eval{e^{igB_{\bm}}e^{-igB_{\bn}}}=\left|\eval{e^{igB}}\right|^2
e^{\frac{g^2}2\nabla^2G(\bm-\bn)}\ ,
\end{equation}
and at large distances (neglecting power-like prefactors),
\begin{equation}
\nabla^2G(x)=-\int\frac{d^2k}{(2\pi)^2}(k^2+m^2)^{1/2}e^{i\bk\cdot\bx}
\sim e^{-mx} \ .
\end{equation}
This dynamically generated mass is Polyakov's result \cite{polyakov}, and is
missing from \cite{DQSW}.

For $\alpha>{\pi^2\over 4}$, (\ref{mass}) has no real solution for $m$.
In this case the energy is minimized at the endpoint of the
variational parameter range, $m^2=0$. 
We conclude that in the Gaussian variational approximation,
mixed action QED$_3$ has a phase transition at $\alpha_c={\pi^2\over 4}$. 
The mass gap vanishes continuously at the critical point while
$(\partial m /\partial\alpha)_{\alpha_c^-}\ne 0$. The phase transition
therefore appears to be second order.\footnote{The 
value of the critical coupling $\alpha_c={\pi^2\over 4}$ should not be taken
too literally, since $G$ is modified from its continuum form close 
to the boundaries of the Brillouin zone. 
It is however
clear that the phase transition is present in any cutoff scheme, at least 
in the present approximation.}

The energy density (\ref{eval}) contains ultraviolet contributions.
The vacuum energy density in (almost) any field theory is ultraviolet
divergent;  thus it should diverge for any variational {\em ansatz} and
indeed (\ref{eval}) contains such a divergence.
The crucial question for us is whether the
leading ultraviolet divergent terms depend
on the variational parameter.  
This would indicate that the parameter is introduced in a way
that directly affects the UV properties of the state.  In other words
such a dependence would mean that one is dealing with a set of states
for which even the UV part of energy is not minimized yet.  Such a
variation would have little to say about IR physics.  

Examining (\ref{eval}) we see that the leading UV divergent terms
are independent of the variational mass $m$. 
The UV behaviour of $G$ is determined (in both phases) independently
of $m$ and is consistent with free field theory in the high momentum limit.
(This indeed should be the case 
since the 3d U(1) gauge theory is superrenormalizable
and its UV physics is governed by a fixed point at zero
coupling.)
Nontrivial
information about the IR physics resides not in the leading terms in
the energy but in the subleading ones.
This makes us confident that the ultraviolet and infrared physics have
indeed been separated in a clean way in our variational calculation.

\subsection{Spacelike Wilson loops}

Let us see whether charges are confined in the two phases.
To this end we calculate the vacuum expectation value of the
spacelike Wilson loop,
\begin{equation}
W_l[C]= \eval{\exp\left(ilg\oint_C \bA\cdot d\bx\right)} = 
\eval{\exp\left(ilg\int_SB\,dS\right)}\ ,
\end{equation}
where $l$ is an integer and the integral is over the area $S$ 
bounded by the loop $C$.
We have
\begin{equation}
W_l[C]=\eval{\prod_S e^{il\pi m_\bn}}Z_A^{-1}\int D\bA \exp\left(
-AG^{-1}A+ilg\int_SB\,dS\right)
\label{wilson}
\end{equation}
The second factor is a Gaussian integral, which gives the factor
\begin{equation}
W_A=\exp\left(\frac{l^2g^2}4\int_Sd^2x\int_Sd^2y\,
\nabla^2G(\bx-\by)\right)\ .
\label{wa}
\end{equation}
In the limit of large $S$ the leading behavior of the exponent is
\begin{equation}
-\frac{l^2}4g^2S\lim_{k\to0}k^2G(k)=-\frac{l^2}{4}g^2mS\ .
\end{equation}
In the massive phase this gives an area law 
with the string tension\footnote{As was shown in \cite{greensite} the 
dependence on $l$ in this formula is incorrect. 
The correct result is $\sigma\propto l$ rather than $\sigma\propto l^2$.
For multiply charged Wilson loops the nonlinearities of the compact theory are
important and the Gaussian variational {\em ansatz\/} may be 
inadequate. This point is peripheral to the present paper and hence we will not 
pursue it.}
\begin{equation}
\sigma=\frac{l^2}{4}g^2m\ .
\label{str}
\end{equation}
In the massless phase this factor does not produce an area law.

The first factor in equation (\ref{wilson}) is different from unity only
for odd $l$.
It can be easily calculated to be
\begin{equation}
W_v\equiv\eval{\prod_S e^{il\pi m_\bn}}=
\int D\chi \exp\left[-g^2\chi D^{-1}\chi+2z\int d^2x
:\!\cos\Big(\chi(x)-\alpha(x)\Big)\!:\right]\ ,
\label{wvort}
\end{equation}
where $\alpha(x)$ vanishes 
for $x$ outside the loop and is equal to $\pi$ for $x$ inside the loop.
Recall that, from (\ref{DG1}),
\begin{equation}
D(k)=4\pi^2k^{-2}G^{-1}(k)=4\pi^2(k^2+m^2)^{-1/2}\ .
\end{equation}
At weak coupling we expand around a classical minimum of the exponent.
In the massive phase, $\alpha<{\pi^2\over 4}$, the inverse propagator
$D^{-1}$ is nonzero at zero momentum and thus it dictates
the leading order solution $\chi(x)=0$. 
For this solution we find
\begin{equation}
W_v=e^{-4zS}\ .
\label{wvortf}
\end{equation}
This is a sub-leading correction to 
the string tension (\ref{str}), since $g^2m\sim e^{-\pi/4g^2}$ and $z\propto m^2
\ll g^2m$.

In the massless phase the situation is different. 
Here the kinetic term of the dual field $\chi$ vanishes at zero momentum, 
$D^{-1}(0)=0$. 
Any constant value of $\chi$ is therefore compatible with the kinetic term.
The classical configuration that minimizes the vortex action in (\ref{wvort})
for large area ($S\gg z^{-1}$) is $\chi=0$ outside the area enclosed by the loop 
and $\chi=\pi$ well inside  the area. 
The field interpolates between these two values in a layer of thickness
$z^{-1}$ in the neighborhood of the loop\footnote{In the
massive phase this classical configuration possesses an enormous action
$\sim g^2mS$, proportional to the area,
since the kinetic term $D^{-1}$ does not vanish at zero momentum.
The leading contribution to the Wilson loop therefore comes from configurations
with $\chi=0$ well inside the loop as stated above.}.
The action for this configuration clearly is proportional to the perimeter.
Thus in the massless phase the string tension vanishes altogether.

The spacelike Wilson loop tells us
that the massive phase is confining with the string tension
related in the expected way to the dynamically generated scale,
$\sigma\propto g^2m$.
The massless phase does not exhibit an area law for spacelike Wilson loops and
we therefore expect that the static charges in this phase are not confined.

\section{The minimally modified state}

As we mentioned in the Introduction, in a Lorentz invariant theory the string
tension extracted from the
spacelike Wilson loop should be the same string tension that determines the
strength of the linear potential between two static charges.
Lorentz symmetry, however, is not maintained in a variational calculation.
Moreover, the theory we are studying is defined
with an explicit ultraviolet regulator that itself breaks Lorentz invariance.
It is thus interesting to perform a separate calculation of the potential 
between external charges.

\subsection{The wave function}

In the charged sector the gauge invariance condition (\ref{g_invariance})
is replaced by
\begin{equation}
\psi[\bA+\nabla\lambda]=\psi[\bA]
\exp\left(ig\sum_\bn \rho_\bn\lambda_\bn\right)\ ,
\label{gauge3}
\end{equation}
where $\rho_\bn$ is a fixed, integer background charge distribution.
(We have in mind a well-separated dipole.)
A wave function that satisfies (\ref{gauge3}) is
\begin{equation}
\psi[\bA]=\sum_{\{m_{\bn'}\}}\int[d\phi_\bn]\,
\exp\left[-\frac12\sum_{\br,\bs}A^{(\phi,m)}_{\br i}
G^{-1}(\br-\bs)A^{(\phi,m)}_{\bs i}\right]
\exp\left(ig\sum_\bn \rho_\bn\phi_\bn\right)\ .
\label{WF1}
\end{equation}
The shifted field in (\ref{WF1}) is defined along the lines of
(\ref{shift2}),
\begin{equation}
A^{(\phi,m)}=A-\nabla\phi-A^S\cdot m\ .
\label{as}
\end{equation}
but now it is necessary to define the vortex field $A^S$ in a singular gauge.
Like $A^V$, the field $A^S$ satisfies
\begin{equation}
\left(\nabla\times\bA^S\right)_{\bn'}=\frac{2\pi}g\delta_{{\bn'},0}\ ,
\end{equation}
but while $A^V$ is divergenceless we take for $A^S$
the solution where $A$ is non-zero on the links dual to
a string extending from 0 (the plaquette of the vortex) to $x=+\infty$,
\begin{eqnarray}
A^S_{\bn x}&=&0\cr
A^S_{\bn y}&=&\left\{
\begin{array}{cl}
\frac{2\pi}g&\text{for}\ n_x>0,\ n_y=0\cr
0&\text{otherwise}
\end{array}
\right.
\end{eqnarray}

Our reason for using $A^S$ rather than $A^V$ in (\ref{WF1})
is one of locality.
A dipole can be created in the vacuum (\ref{wf}) by the string operator
$\exp\left[ig\sum_{\bx}^{\by}A(\bz)\right]$, which places sources at
\bx\ and \by.
This operator creates a string of electric flux taking integer values
along the string, the most local way of preserving Gauss' Law when creating
a dipole.
A shift of \bA\ by $A^S$ commutes with this string operator, but a shift by
$A^V$ does not.
Shifting by $A^V$ will create a non-local, transverse electric field with
{\em fractional} flux in addition to the string. In this light it
appears unavoidable that the introduction of dynamical charges will
immediately lead to a nonlocal and non-Lorentz-invariant theory.
For more on this point see \cite{DQSW}.

Another way of looking at it is to note that the difference between
$A^V$ and $A^S$ can be absorbed by a shift in the integration variable
$\phi_{\br}$ by ${1\over g}\theta(\br-\br')m_{\br'}$,
where $\theta(\br-\br')$ is the angle that the vector
$\br-\br'$ makes with the $x$ axis.
Using $A^V$ would therefore lead to an extra phase factor 
$\exp\left[i\sum_{\br,\br'} \rho_\br \theta(\br-\br')m_{\br'}\right]$
in the integral in (\ref{WF1}).
Under a shift of vortex density $m$ this
function is not invariant, but rather acquires a phase
proportional to the charge density. The shift of vortex density can be
viewed as a kind of large gauge transformation \cite{KK}
and therefore this wave function belongs to a sector of the Hilbert
space with a position-dependent
``$\theta$ angle.'' 
It is hard to imagine how this sector can define a
local theory, especially in the presence of dynamical charges.

Since $A^S$ differs from $A^V$ by a gauge transformation, 
it can be used interchangeably with it in the vacuum wave function (\ref{wf}).
The distinction is only meaningful in the charged sector, where it leads to
a new vortex--charge interaction [see (\ref{part1}) below].

Taking $G$ to be the same function as was used in Section \ref{sec2}
in the vacuum sector,
this is the {\em minimally modified state\/} used by \cite{DQSW,zare,dyak}.

\subsection{Normalization and expectation values}

The normalization factor for the wave function (\ref{WF1}) is
\begin{equation}
Z=Z_AZ_\phi[\rho] Z_v [\rho]
\end{equation}
with
\begin{eqnarray}
Z_A&=&\det \pi G\ ,\\ 
Z_\phi[\rho]&=&
\left(\det 4\pi\frac1{\nabla^2}G\right)^{1/2}
\exp\left(-g^2\rho\cdot\frac1{\nabla^2}G\cdot\rho\right)\ ,\\
Z_v[\rho]&=&\sum_{\{m_{\bn'}\}}\exp\left[-\frac1{4g^2}\sum_{\br',\bs'}m_{\br'}
D({\br'}-{\bs'})m_{\bs'} \right]
\exp\left(-i\sum_{\br'}h_{\br'}m_{\br'}\right)\ .
\label{part1}
\end{eqnarray}
The new ingredient in (\ref{part1}) is a vortex--charge interaction,
\begin{equation}
\sum_{\br'}h_{\br'}m_{\br'}\equiv
\sum_{\br\br'}\rho_{\br}\theta(\br-\br')m_{\br'}\ .
\end{equation}
The vortex interaction potential $D$ is again given by
(\ref{DG1}).

First let us calculate the expectation value of the electric field in
this state. Straightforward algebra gives
\begin{equation}
\eval{E_i}=\eval{i\frac{\partial}{\partial A_i}}=
g\frac{\partial_i}{\nabla^2}\rho-
\frac{i\pi}{g}G^{-1}
\frac{\epsilon_{ij}\partial_j}{\nabla^2}\eval{m}
\label{field}
\end{equation}
The first term on the right hand side is the longitudinal field
that gives the Coulomb interaction between the charges in non-compact
electrodynamics. 
The second term,
entirely due to the compactness of the theory,
is purely transverse as it must be in order not to spoil 
Gauss' law.
 
The calculation of the energy follows the same lines as the analogous 
calculation discussed in some detail in the previous section.
For the kinetic term [cf. (\ref{elen1})],
\begin{eqnarray}
\eval{\int E^2\,d^2x}&=&\frac{V}2\int\frac{d^2k}{(2\pi)^{2}}G^{-1}(k)-
g^2\int d^2x\,\rho\frac1{\nabla^2}\rho\nonumber\\
&&\qquad
+{\pi^2\over g^2}\int d^2x\,d^2y\,\partial^{-2}G^{-2}(x-y)\eval{m(x)m(y)}
\end{eqnarray}
while the potential term is
\begin{equation}
\eval{-\frac1{g^2}\left(\alpha \cos gB+\frac{1-\alpha}4 \cos2gB\right)}=
\frac14\int {d^2k\over (2\pi)^2} k^2G(k)
-\frac{\alpha}{g^2}\left(\text{Re}\eval{e^{i\pi m}}-1\right)\ .
\end{equation}
Putting these together we obtain for the energy of the minimally modified state
\begin{eqnarray}
\eval{H}&=&\frac V4\int\frac{d^2k}{(2\pi)^2}
\left[G^{-1}(k)+k^2G(k)\right]
-\frac{g^2}2\int d^2x\,\rho\frac1{\nabla^2}\rho\nonumber\\
&&\qquad
+\frac{\pi^2}{2g^2}\int d^2x\,d^2y\,\partial^{-2}G^{-2}(x-y)\eval{m(x)m(y)}
\nonumber\\&&\qquad
-\frac{\alpha}{g^2}\int d^2x\,\left(\text{Re}\eval{e^{i\pi m(x)}}-1\right)\ .
\label{energy1}
\end{eqnarray}

To calculate correlation functions of the vorticity $m$,
we again introduce the dual field $\chi$ as in (\ref{duality}) and
obtain the relations
\begin{eqnarray}
\eval{m}&=&-2ig^2D^{-1}\eval{\chi}\nonumber\\
\eval{m(x)m(y)}&=&2g^2D^{-1}-4g^4\eval{D^{-1}\chi(x) D^{-1}\chi(y)}\ ,
\label{rhochi}
\end{eqnarray}
where the averages of the dual field $\chi$ are calculated with the
partition function
\begin{equation}
Z_v=\int D\chi\exp\left[-g^2\chi D^{-1}\chi+
2z\int d^2x:\!\cos\Big(\chi(x) -\theta\rho\Big)\!:\right]\ .
\label{sine1}
\end{equation}
[The notation $\theta\rho$ represents the convolution of the source
distribution $\rho(\br)$ with $\theta(\br-\br')$.]
In the weak coupling (small $z$) regime
we solve the classical equation of motion,
\begin{equation}
g^2D^{-1}\chi+z\sin(\chi-\theta\rho)=0
\label{class}\end{equation}
We consider the massive ($\alpha<\alpha_c$) and massless
($\alpha>\alpha_c$) cases separately.

\subsection{Field profile and energy in the massive phase}

First let us take $\alpha<\alpha_c$. 
In this case the variational propagator $G$
is massive.
This mass is large compared to $z$ since $z\propto m^2\ll g^2m$.
Recall that $D^{-1}\sim \sqrt{k^2+m^2}$.
The classical equation (\ref{class}) can then
be solved in perturbation theory in $\frac{z}{g^2m}$.
To first order in $z$ we find
\begin{equation}
\chi=\frac{zD}{g^2}\sin(\theta\rho)
\label{chim}
\end{equation}
and thus
\begin{eqnarray}
\eval{m(x)}&=&-2iz\sin(\theta\rho(x))\nonumber\\
\eval{m(x)m(y)}_{\text{connected}}&=&2z\cos(\theta\rho(x))\delta^2(x-y)\\
\eval{e^{i\pi m(x)}}&=&\exp[-4z\cos(\theta\rho(x))]\nonumber
\end{eqnarray}

Our first observation is that with this average value of the vortex density
the electric field far away from the charges is not screened, but rather
decays the same way as in the non-compact theory.
To see this let us rewrite the expression for the electric field in terms of the
dual field $\chi$,
\begin{equation}
\eval{E_i}=g\frac{\partial_i}{\nabla^2}\rho+
\frac{g}{2\pi}\epsilon_{ij}\partial_j\eval\chi\ .
\end{equation}
In order for the electric field to be screened away from the charges,
the first and second terms must cancel. 
For this to happen the dual field $\chi$ must behave asymptotically as
\begin{equation}
\chi(x)\to\theta\rho(x)\text{ as }x\to\infty\ .
\label{chican}
\end{equation}
To find the large-distance behavior of the solution (\ref{chim})
we take for $D(k^2)$ its
value at zero momentum, $D(0)=(4\pi^2m)^{-1}$. We then find
\begin{equation}
\chi(x\rightarrow\infty)\sim\frac{z}{g^2m}\sin(\theta\rho(x))
\label{chi1}
\end{equation}
which is much smaller than required by (\ref{chican}), so
there is {\em no} cancellation of the long range part of the electric field.

Technically it is clear why we get the estimate (\ref{chi1}) rather
than (\ref{chican}) for the asymptotic behavior of the classical
solution. 
If we were to
minimize only the potential term in the dual action, (\ref{chican}) would be 
the result.
In the massive phase, however, the
kinetic term $\chi D^{-1}\chi$ is the leading term for
configurations dominated by low momentum components and it
determines the infrared behavior of the solution.
To illustrate this let
us estimate the dual field action (\ref{sine1}) for a field
configuration that behaves as (\ref{chican}).
We take $\rho$ to be a charge dipole with separation $L$.
The main contribution to the action comes from the kinetic term in the 
infrared region,
\begin{equation}
S_\chi\sim g^2mL^2\log V\ ,
\label{s1}
\end{equation}
where $V$ is the total volume (area) of the system.
For the function (\ref{chim}) on the other hand the contributions
of the kinetic and potential terms are of the same order and turn out
to be
\begin{equation}
S_\chi\sim zL^2\log V\ .
\end{equation}
This is much smaller than
(\ref{s1}) and therefore the configuration (\ref{chim})
dominates the path integral over $\chi$.

The electric field of a pair of static charges thus does not decay
exponentially
in the minimally modified state, even though the vacuum does contain a
dynamically generated scale.
This leads us to suspect that the the minimally modified state
is far from being the ground state in this sector.
A calculation of the expectation value of the energy confirms this suspicion.
The crucial terms are  the last two terms, the vortex terms, 
in (\ref{energy1}). 
To first order in $z$ the contribution of this term to the energy of the 
minimally modified state relative to the vacuum is
\begin{equation}
\Delta E=m^2g^2\int d^2x\left[1-\cos(\theta\rho)\right]
\label{delta}
\end{equation}
Far from the charges, we have $\theta\rho(x)\rightarrow L{x_1/x^2}$
when the dipole is parallel to the $x_1$ axis.
The integral in (\ref{delta}) is positive and
{\em logarithmically divergent in the volume}.
It is hard to believe that this infrared divergent energy has anything
to do with the actual interaction energy of the charges.

\subsection{The massless phase}

In the massless phase, the kinetic term of the dual field vanishes at
zero momentum, $D^{-1}(k^2)\sim |k|$.
The solution therefore can not be expanded in powers of $zD$ as before.
The infrared behavior of the solution, however, can still be easily understood.
It is determined by the potential term in (\ref{sine1}) and is given by
(\ref{chican}). 
The distance at which this behavior sets in is
determined by the value of momentum at which the kinetic and potential
terms are comparable, $x^{-1}\propto|k|={z}/{g^2}$.
The electric field is screened away from the dipole, decaying\footnote{The
screening here is not exponential but rather power-wise. This is due to
the nonanalyticity of the $\chi$ propagator at zero momentum, $z\not=0$
notwithstanding.
Expanding $D^{-1}$ around $k=0$ we find that the leading term $D^{-1}\sim|k|$, 
gives in coordinate space $\eval{\chi\chi}\sim1/x^3$.}
faster than $1/x$.
The ``flux tube,'' carrying flux $g$ in a region of transverse size $g^2/z$,
has energy
\begin{equation}
\Delta E\sim zL\ .
\end{equation}
This is precisely the result of \cite{DQSW}.
By itself it would suggest that the external charges are 
confined with the string tension $\sigma=z$, despite the absence of an area
law for the spacelike Wilson loop. 
We suspect, however,
that just as in the massive phase, this contradiction
indicates that the minimally modified state provides a very poor bound on the
energy of the charges.

A more reliable calculation of the interaction energy
within the variational framework requires
the introduction of additional variational parameters. This we do 
in the next section.

\section{The interaction energy in the extended variational ansatz}

As we have seen in the previous section the minimally 
modified state has a prescribed profile of the electric field.
This profile, as it turns out, does not conform to one's 
intuitive picture of the profile of the electric field in the charged state. 
When there is a dynamically generated mass and an area
law for the spacelike Wilson loop, one expects the electric field
between the external charges to form a flux tube.
In the minimally modified state 
the field is just the Coulomb field of an electric dipole. 
In the massless phase, conversely, we expect to find a dipole field, but 
instead we obtain a screened distribution similar to a flux tube.
Moreover, the energy of the minimally modified state is unnaturally high.

\subsection{The extended ansatz}

The above leads us to believe that the minimally modified state is very far in
its physical properties from the minimal energy state in the charged sector.
It is apparent that the main missing feature is freedom  
for the electric field to adjust its profile. 
We therefore introduce this additional variational freedom in our {\em ansatz},
writing
\begin{eqnarray}
\psi[\bA]&=&\sum_{\{m_{\bn'}\}}\int[d\phi_\bn]\,
\exp\left[-\frac12\sum_{\br,\bs}A^{(\phi,m)}_{\br i}
G^{-1}(\br-\bs)A^{(\phi,m)}_{\bs i}\right]
\exp\left(ig\sum_\bn\rho_\bn\phi_\bn\right)\cr
&&\times\exp\left(i\sum_\bn{\bf e}_\bn\cdot\bA^{(\phi,m)}
_\bn\right)\ .
\label{wf2}
\end{eqnarray}
We take the classical background field
$\be$ to be transverse, $\nabla\cdot\be=0$, and we will
treat it as an additional variational parameter in the charged
sector.

Retracing the calculations of the previous sections 
we find that the normalization $Z$ is modified only in the vortex
factor (\ref{part1}), wherein the external field $h_{\br'}$ acting on
the vortices becomes
\begin{equation}
h=\frac{2\pi}g\frac1{\nabla^2}\epsilon_{ij}\partial_i e_j
+\theta\rho\ .
\label{hfield}
\end{equation}
The expectation value of the electric field is
\begin{equation}
\eval{E_i}=
g\frac{\partial_i}{\nabla^2}\rho-
\frac{i\pi}{g}G^{-1}
\frac{\epsilon_{ij}\partial_j}{\nabla^2}\eval{m}
+e_i\ .
\label{field1}
\end{equation}
[Note that $\be$ enters (\ref{field1}) as a simple shift and also as part of
the external field (\ref{hfield}) that determines $\eval m$.]
The energy of this state is
\begin{eqnarray}
\eval{H}&=&\frac V4\int\frac{d^2k}{(2\pi)^2}
\left[G^{-1}(k)+k^2G(k)\right]
-\frac{g^2}2\int d^2x\,\rho\frac1{\nabla^2}\rho\nonumber\\
&&\qquad
+\frac{\pi^2}{2g^2}\int d^2x\,d^2y\,\partial^{-2}G^{-2}(x-y)\eval{m(x)m(y)}
\nonumber\\&&\qquad
-\frac{\alpha}{g^2}\int d^2x\,\left(\text{Re}\eval{e^{i\pi m(x)}}-1\right)
\nonumber\\&&\qquad
+\frac{i}{2}\zeta G^{-1} \eval{m}
+\frac12\int d^2x\,e^2\ ,
\label{energy2}
\end{eqnarray}
where the penultimate term contains the potential $\zeta(x)$, defined via
\begin{equation}
e_i=\frac g{2\pi}\epsilon_{ij}\partial_j\zeta\ .
\label{zeta}
\end{equation}
Thus we have also
\begin{equation}
h=-\zeta+\theta\rho\ .
\end{equation}
The duality transformation now results in the following Lagrangian for the
dual field $\chi$,
\begin{equation}
L=g^2\chi D^{-1} \chi-
2z\int d^2x\,:\!\cos\Big(\chi(x)+\zeta -\theta\rho\Big)\!:\ .
\end{equation}
The correlation functions of $m$ are related to the correlation 
functions of $\chi$ via (\ref{rhochi}).

\subsection{The massive phase}

In the massive phase, we obtain as before to first order in $z$ 
\begin{eqnarray}
\eval{m(x)}&=&-2iz\sin(\theta\rho(x)-\zeta(x))\nonumber\\
\eval{m(x)m(y)}&=&2z\cos(\theta\rho(x)-\zeta(x))\delta^2(x-y)
\label{rho2}\\
\eval{e^{i\pi m(x)}}&=&\exp[-4z\cos(\theta\rho(x)-\zeta(x))]\nonumber
\end{eqnarray}
The $\be$-dependent piece of the energy is
\begin{eqnarray}
\Delta E&=&m^2g^2\int d^2x\,[1-\cos(\theta\rho(x)-\zeta(x))]\nonumber\\
&&\quad+z\int d^2x\,d^2y\,\zeta(x) 
G^{-1}(x-y)\sin(\theta\rho(y)-\zeta(y))\nonumber\\
&&\quad-\frac{g^2}{8\pi^2}\int d^2x\,\zeta\nabla^2\zeta
\label{en}
\end{eqnarray}
The quantity (\ref{en}) is to be minimized with respect to $\zeta$. 
It is obvious immediately that we will have $\zeta\to\theta\rho$ at large
distances from the sources, so that the infrared divergence found in the
minimally modified state will disappear.
Noting that the second term is formally of order $g^2$ relative to the
first, we drop it for now, subject to a consistency check at the end.
The minimization equation for $\zeta$ then becomes very simple,
\begin{equation}
{\nabla^2\zeta-m^2\sin(\zeta-\theta\rho)=0\ .}
\label{sgmin}
\end{equation}

To study (\ref{sgmin}) it is convenient to define
$\tilde\zeta=\zeta-\theta\rho$ which satisfies the sine-Gordon equation
with a singular source term,
\begin{equation}
{\nabla^2\tilde\zeta-m^2\sin\tilde\zeta=S}
\label{sineg}
\end{equation}
The source term $S$ consists of a dipole layer along the line between
the external point charges.
For charges separated by a distance $L$, we have
\begin{equation}
S(\bx)=2\pi\,\delta'(x_2)\,\eta\left(\frac L2+x_1\right)\,
\eta\left(\frac L2-x_1\right)\ ,
\end{equation}
where $\eta$ is a step function.
When $L$ is much larger than $1/gm$ the solution
of (\ref{sineg}) can be found in the region $-L/2\ll x_1\ll L/2$.
In this region $\tilde\zeta$
is approximately independent of $x_1$ and satisfies
the one-dimensional sine-Gordon equation in $x_2$,
\begin{equation}
{\frac{\partial^2\tilde\zeta}{\partial x^2_2}-m^2\sin\tilde\zeta=S\ .}
\end{equation}
The singular source $S$ creates a discontinuity of $2\pi$ at $x_2=0$.
The solution for $x_2>0$ is hence half a sine-Gordon soliton, with
$\tilde\zeta=\pi$ at $x_2=0+$ and $\tilde\zeta\to0$ as $x_2\rightarrow\infty$.
For $x_2<0$ it is half the antisoliton with $\tilde\zeta=-\pi$ at $x_2=0-$ and
$\tilde\zeta\to0$ as $x_2\to-\infty$.
At distances greater than $1/m$ from the $x_1$ axis, $\tilde\zeta$ vanishes 
exponentially.
Therefore we have $\zeta\to\theta\rho$ {\em exponentially}.
The vortex density $m(x)$ according to (\ref{rho2}) {\em vanishes}
exponentially.
Referring to (\ref{field1}) we see that the total electric
field vanishes exponentially outside the region of width $1/m$, indeed
forming a flux tube of thickness $1/m$. 
This is perfectly in accord with our expectation for the field profile in a 
confining theory.

In (\ref{en})
we see that the energy density of the flux tube is proportional to the energy 
of the sine-Gordon soliton solution with the proportionality 
coefficient $g^2\over 4\pi^2$.  
The string tension therefore is
\begin{equation}
\sigma={2\over\pi^2}\,g^2m\ .
\end{equation}
This is again consistent with the result of the calculation of the
spacelike Wilson loop in Section 2.
We do get a slightly lower value for the string tension (a factor of
$2/\pi^2$ rather than $1/4$) here than we do from the spatial Wilson loop.
This is a consequence of minimizing the
energy independently in the charged sector.

\subsection{The massless phase}

{The crucial feature of the massless phase is that now $D^{-1}$
vanishes at zero momentum. 
It is therefore incorrect to expand the
classical solution for the dual equation of motion in powers of $zD$.
Let us define $\chi_c$ to be the
solution of the classical equation of motion for the dual field,
\begin{equation}
g^2D^{-1}\chi_c+z\sin(\chi_c+\zeta-\theta\rho)=0\ .
\label{chic}
\end{equation}
For our purposes it will not be necessary to know the form of
$\chi_c$ explicitly.
The essential point is that unlike in the massive phase, for the 
$\chi$ propagator we must use the complete expression
\begin{equation}
\eval{\chi\chi}_{\text{connected}}=\left[2g^2D^{-1}+2z\cos(\chi_c+\zeta
-\theta\rho)\right]^{-1}
\end{equation}
{\em without\/} expanding in $z$.
Calculating the correlation functions of the vorticity via (\ref{rhochi}), and
substituting into (\ref{energy2}), we obtain for the energy
\begin{eqnarray}
&&\Delta E=
-{g^2\over 8\pi^2}(\chi_c+\zeta)\nabla^2(\chi_c+\zeta)\label{en1}\\
&&\quad+4g^4\left({\pi^2\over 2g^2}-{2\alpha\over g^2}\right)
{\rm Tr} \,D^{-1}\left\{\left
[2g^2D^{-1}+2z\cos(\chi_c+\zeta-\theta\rho)\right]^{-1}
-\left[2g^2D^{-1}+2z\right]^{-1}\right\}D^{-1}\ .\nonumber
\end{eqnarray}
To find the infrared asymptotics we neglect $D^{-1}$ in the
denominators in (\ref{en1}) since it vanishes at zero momentum.
The second term then becomes 
\begin{equation}
\int d^2x\,\nabla^2 \left({1\over \cos(\chi_c+\zeta-\theta\rho)}\right)\ .
\end{equation}
This is the integral of a total derivative and it vanishes since 
$1/\cos(\chi_c+\zeta-\theta\rho)$ does not diverge at infinity.}

The profile of $\zeta$ is therefore determined by minimizing the first term in
(\ref{en1}).
The solution is
\begin{equation}
\chi_c+\zeta=0\ .
\end{equation}
Thus at distances larger than $g^2/z$,
the induced electric field ${g\over 2\pi}\epsilon_{ij}\partial_j\zeta$ 
cancels the contribution of the
vortices ${g\over 2\pi}\epsilon_{ij}\partial_j\chi_c$. 
The total electric field at large distances is therefore just
the field of the electric dipole. 
This is again consistent with our
expectation for the behavior of electric field in a massless phase.

\section{Discussion}
In this paper we studied compact QED$_3$ with mixed action in the
gauge invariant Gaussian variational approach. We find that the theory
has two phases. When $\beta$, the coefficient of the charge-2 term in the 
action, is not too negative the theory is massive and confining. 
For $\beta$ large and negative no mass is generated and the
theory is not confining. 
We have also calculated variationally the interaction energy
of a state with two external charges.

The authors of \cite{DQSW} studied a minimally modified variational state
with {\em no\/} dynamical mass within the framework of the pure fundamental
theory. 
As emerges from our analysis the vacuum of the pure fundamental theory 
is massive. This mass has to be introduced explicitly in the Gaussian. 
The nonvanishing string tension is due to this dynamical mass and behaves
according to the Polyakov formula $\sigma\sim g^2m\sim z^{1/2}$ where
$z$ is the vortex fugacity.

The massless state is sensible in the theory with mixed
action for $\alpha>\pi^2/4$. 
This phase is not confining. 
The calculation of \cite{DQSW}, however, yields a
string tension $\sigma\sim z$ which is nonvanishing, though parametrically 
smaller than in the massive phase. 
We now understand that this nonvanishing string tension is
an artifact of the minimally modified state.  
A better variational state does not exhibit flux tube formation
in the massless phase.

The main lesson we have learned is that to get a good estimate for the
interaction energy of external charges
we have to introduce a background electric field as an
additional variational parameter. 
The minimally modified state of \cite{zare,dyak} which does not allow for
this variation is quite poor energetically. 
It has a wrong profile of the electric field and as a
consequence  overestimates significantly the interaction energy.
We expect the same to be true in nonabelian gauge theories, so we expect
the calculation of the interaction energy of external charges there to be more
complicated than envisioned in \cite{zare,dyak}.

On the positive side we found that the calculation of the 
spacelike Wilson loop gives a good estimate for the interaction
potential and the associated string tension even though the
variational wave function, and indeed the cutoff Hamiltonian theory,
is not Lorentz invariant.

\section*{Acknowledgments}
A.~K. thanks the Department of Nuclear Physics at Tel Aviv University
for their hospitality.
B.~S. thanks the 
Theoretical Physics group at Oxford University for hospitality
and PPARC for financial support during his visit there.
He also thanks the Weizmann Institute of Science for its continuing
hospitality.
A.~K. is supported by a PPARC Advanced Fellowship.
The work of B.~S. is partially supported by the Israel Science Foundation
under Grant~No.~255/96-1.


\begin{thebibliography}{99}

\bibitem{polyakov} A. M. Polyakov,  Phys. Lett. {\bf 59B}, 82 (1975);
Nucl. Phys. B {\bf 120}, 429 (1977).

\bibitem{CDG}C.~G.~Callan, R.~Dashen and D.~J.~Gross,
Phys. Rev. D {\bf 17}, 2717 (1978); {\em ibid.} {\bf 19}, 1826 (1979).

\bibitem{BMK}T. Banks, R. Myerson and J. Kogut, Nucl. Phys. B {\bf 129},
493 (1977).

\bibitem{Creutz}M.~Creutz, Phys. Rev. D {\bf 21}, 2308 (1980).

\bibitem{DQSW} S. D. Drell, H. R. Quinn, B. Svetitsky, and M. Weinstein,
Phys. Rev. D {\bf 19}, 619 (1979).

\bibitem{koko1}I. Kogan and A. Kovner, Phys. Rev. D {\bf 52},
3719 (1995).

\bibitem{iancu} C. Heinemann, C. Martin, D. Vautherin, and E. Iancu,
hep-ph/9802036.

\bibitem{KK} I. Kogan and A. Kovner, Phys. Rev. D {\bf 51}, 1948 (1995).

\bibitem{brko} W. E. Brown and I. I. Kogan, hep-th/9705136.

\bibitem{brown} W. E. Brown, hep-th/9711189. 

\bibitem{zare} K. Zarembo, Phys. Lett. B {\bf 421}, 325 (1998);
hep-th/9806150.

\bibitem{dyak} D. Diakonov, hep-th/9805137.

\bibitem{Morris}T. R. Morris, Phys. Rev. D {\bf 53}, 7250 (1996).

\bibitem{Shahar}S. Ben-Menahem, Phys. Rev. D {\bf 20}, 1923 (1979).

\bibitem{greensite} J. Ambjorn and J. Greensite, J. High Energy Phys. {\bf 05}
(1998) 004, hep-lat/9804022.


\end{thebibliography}
\end{document}